\documentclass[twocolumn]{aa}
\usepackage{graphicx}
\newcommand{\psim}{\lower.5ex\hbox{$\; \buildrel \propto \over\sim \;$}}
\newcommand{\lesssim}{\lower.5ex\hbox{$\; \buildrel < \over\sim \;$}}
\newcommand{\gtrsim}{\lower.5ex\hbox{$\; \buildrel > \over\sim \;$}}

\newcommand{\e}{\epsilon}

\newcommand{\lC}{\bar\lambda_{\rm C}}

\begin{document}
   \title{Neutral beam model for the anomalous $\gamma$-ray emission 
component in GRB 941017}


   \author{C.\ D.\ Dermer          \inst{1}
          \and
          A.\ Atoyan\inst{2}
          }

   \offprints{C.\ D.\ Dermer}

   \institute{Code 7653, Naval Research Laboratory, 
Washington, DC 20375-5352 USA\\
              \email{dermer@gamma.nrl.navy.mil,
 http://heseweb.nrl.navy.mil/gamma/$\sim$dermer/default.htm}
         \and
             Centre de Recherches Math\'ematiques, 
Universit\'e de Montr\'eal, Montr\'eal, Canada H3C 3J7\\
             \email{atoyan@crm.umontreal.ca}
             }

   \date{Received January 5, 2004; accepted March 4, 2004}
\authorrunning{Dermer \& Atoyan}
\titlerunning{Neutral Beam Model for GRB 941017}

\abstract{
Gonz\'alez {\it et~al}.\ (2003) have reported the discovery of an
anomalous radiation component from $\approx 1$ -- 200 MeV in GRB
941017.  This component varies independently of and contains $\gtrsim
3\times$ the energy found in the prompt $\sim 50$ keV -- 1 MeV
radiation component that is well described by the relativistic
synchrotron-shock model. Acceleration of hadrons to very high energies
can give rise to two additional emission components, one produced
inside the GRB blast wave and one associated with an escaping beam of
ultra-high energy (UHE; $\gtrsim 10^{14}$ eV) neutrons, $\gamma$ rays,
and neutrinos. The first component extending to $\sim 100$ MeV is from
a pair-photon cascade induced by photomeson processes with the
internal synchrotron photons coincident with the prompt radiation.
The outflowing UHE neutral beam can undergo further interactions with
external photons from the backscattered photon field to produce a beam
of hyper-relativistic electrons that lose most of their energy during
a fraction of a gyroperiod in the assumed Gauss-strength magnetic
fields of the circumburst medium. The synchrotron radiation of these
electrons has a spectrum with $\nu F_\nu$ index equal to $+1$ that can
explain the anomalous component in GRB 941017. This interpretation of
the spectrum of GRB 941017 requires a high baryon load of the
accelerated particles in GRB blast waves. It implies that most of the
radiation associated with the anomalous component is released at
$\gtrsim 500$ MeV, suitable for observations with {\it GLAST}, and
with a comparable energy fluence in $\sim 100$ TeV neutrinos that
could be detected with a km-scale neutrino telescope like IceCube.
\keywords{gamma ray bursts -- cosmic rays -- radiation processes } 
}

   \maketitle
%

\section{Introduction}

Based on joint analysis of BATSE LAD (Large Area Detector) and the
EGRET TASC (Total Absorption Shower Counter) data, Gonz\'alez et al.\
(2003) recently reported the detection of an anomalous MeV emission
component in the spectrum of GRB 941017 that decays more slowly than
the prompt emission detected with the LAD in the $\approx 50$ keV -- 1
MeV range. The multi-MeV component lasts for $\gtrsim 200$ seconds
(the $t_{90}$ duration of the lower-energy prompt component is 77
sec), and is detected with the BATSE LAD near 1 MeV and with the EGRET
TASC between $\approx 1$ and 200 MeV. The spectrum is very hard, with
a photon number flux $\phi(\e)\propto \e^{-1}$, where $\e =
h\nu/m_ec^2$ is the observed dimensionless photon energy.

This component is not predicted or easily explained within the
standard leptonic model for GRB blast waves, though it possibly could
be related to Comptonization of reverse-shock emission by the forward
shock electrons (\cite{gg03}), including self-absorbed reverse-shock
optical synchrotron radiation (\cite{pw04}). Another possibility is
that hadronic acceleration in GRB blast waves could be responsible for
this component.

We propose a model involving acceleration of hadrons at the
relativistic shocks of GRBs. A pair-photon cascade initiated
by photohadronic processes between high-energy hadrons accelerated in
the GRB blast wave and the internal synchrotron radiation field
produces an emission component that appears during the prompt
phase. Photomeson interactions in the relativistic blast wave also produce
a beam of UHE neutrons, as proposed for blazar jets
(\cite{ad03}). Subsequent photopion production of these neutrons with
photons outside the blast wave produces a directed hyper-relativistic
electron-positron beam in the process of charged pion decay and the
conversion of high-energy photons from $\pi^0$ decay. These energetic
leptons produce a synchrotron spectrum in the radiation
reaction-limited regime extending to $\gtrsim$ GeV energies, with
properties in the 1 -- 200 MeV range similar to that measured from GRB
941017. If our model is correct, detection of this component
therefore gives important evidence for the acceleration of
UHE cosmic rays in GRB blast waves, and would be confirmed
by the detection of UHE neutrinos from GRBs.

\section{ Hadronic Model for GRB 941017}

We assume that efficient proton acceleration to ultra-relativistic
energies takes place in GRB blast waves, as required in models where
GRBs accelerate high-energy cosmic rays (\cite{vie95,wax95,der02}).
Photopion interactions of these protons with internal synchrotron
photons and with photons from an external radiation field create
neutral particles and charged pions that decay and initiate an
electromagnetic cascade within the GRB blast wave.

During the prompt emission phase, an additional component of radiation
is produced by the cascade occurring within the blast wave.  Fig.\ 1
shows the hadron-initiated cascade radiation for a model GRB at
redshift $z =1 $, with hard X-ray fluence $\Phi_{tot} = 3\times
10^{-4}\,\rm erg \; cm^{-2}$, a light curve of 100 second duration
divided into 50 pulses of 1 second each, and Doppler factor $\delta =
100$ (see \cite{ad03,da03} for more details about the model). The
total amount of accelerated proton energy $E^\prime = 4\pi d_L^2
\Phi_{tot} \delta^{-3} (1+z)^{-1}$ is injected into the comoving frame
of the GRB blast wave.

\begin{figure}[t]
\vskip-0.6in
\vspace*{15.0mm} %
\includegraphics[width=8.3cm]{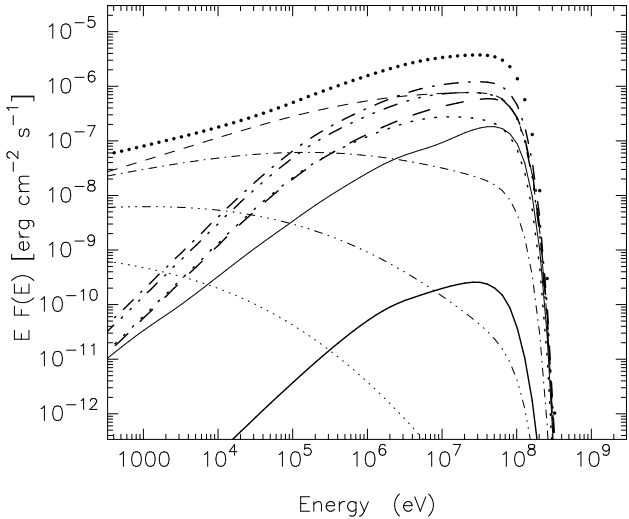}
\caption{Photon energy fluence from an electromagnetic cascade initiated
by photopion secondaries in a model GRB, with parameters given in the
text. Five generations of Compton (heavy curves) and synchrotron
(light curves) are shown. The first through fifth generations are
given by solid, dashed, dot-dashed, dot-triple--dashed, and dotted
curves, respectively. The total cascade radiation spectrum is given by
the upper bold dotted curve.  }
\label{cascade}
\end{figure}

The various generations of synchrotron and Compton radiation initiated
by the cascade are shown in Fig.\ 1, along with the total radiation
spectrum. As can be seen, the isotropic cascade radiation approaches
the spectrum of an electron distribution cooling by synchrotron
losses, that is, a photon number index between $-1.5$ and $-2$. During 
intervals (a) and (b) of GRB 941017 when the prompt hard X-ray
and soft $\gamma$-ray flux is brightest and the high-energy spectral
index is hard though not well-measured (\cite{gon03}), this component could
make the dominant contribution to $\gtrsim 1$ MeV radiation fluence. 
The isotropic cascade radiation will decay at the
same rate as the synchrotron radiation, after which a second 
component from the ouflowing neutral beam begins to dominate.

Ultra-relativistic neutrons formed in the reaction
$p+\gamma\rightarrow n +\pi^0$ are not confined by the magnetic field
in the GRB blastwave shell and flow out to create an energetic neutron
beam. These neutrons are subject to further photopion processes with
photons in the surrounding medium to form charged and neutral pions.
In the Gauss-type magnetic fields surrounding GRB sources that we
assume here, charged $\pi$ and $\mu$ at energies $\lesssim 10^{18}$ eV
decay rather than lose energy through synchrotron emission
(\cite{rm98}). The charged pions decay into ultrarelativistic
electrons and neutrinos, whereas the decay of $\pi^0$ produces two
$\gamma$ rays that are promptly converted into electron-positron pairs
on this same external radiation field. These energetic electrons
(including positrons) are initially produced in the direction of the
GRB jet.

The spectra of secondary electrons created by the neutron beam
displays a sharp cutoff at energies $\lesssim 10^{14}$ eV as a
consequence of the high threshold for photomeson interactions 
(see calculations in \cite{da03}). Electrons with Lorentz factor
$\gamma$ lose energy through synchrotron radiation in an ordered
magnetic field with strength $B$ at the rate $-d\gamma/dt =
\sigma_{\rm T} B^2 \gamma^2\sin^2\psi/(4\pi m_e c)$, where $\psi$
is the electron pitch angle. The corresponding synchrotron energy-loss
time scale $t_{syn} = \gamma/|d\gamma/dt|$. The gyration frequency
$\omega_B = eB/\gamma m_e c $, and is independent of pitch angle. When
$\omega_B t_{syn} \ll 1$, the electron loses almost all of its energy
into synchrotron radiation in a time less than the gyroperiod.  We use
the term ``hyper-relativistic'' to refer to electrons in this
radiation-reaction regime of synchrotron emission (\cite{nw91,ab02}).

Electrons which cool before being deflected by an angle $\theta$ equal
to the jet opening angle $\theta_j$ will emit most of their energy
within $\theta_j$. The pitch angle $\psi$ does not change in the
process of synchrotron losses when $\gamma \gg 1$. Then $\cos\theta =
\cos^2\psi+\sin^2\psi\cos\phi$, where $\phi = \omega_B t$ is the 
rotation angle. In the limit of small $\theta$ and $\phi$, $\theta
\cong \phi \sin\psi$. The condition $\theta \leq \theta_j$ for times
$t \leq t_{syn}$ then results in the condition $\gamma
 \gtrsim \gamma_{hr}(\theta_j) = 
\sqrt{4\pi e/(\theta_j \sigma_{\rm T} B \sin\psi )}$ $\cong 3\times 10^8/
[\sqrt{(\theta_j/0.1) B({\rm G})\sin\psi}]$ for the hyper-relativistic
electrons.  Here we have taken a typical jet opening half-angle
$\theta_j = 0.1$ because this is the average value implied by the
analysis leading to the standard energy reservoir result of GRBs
(\cite{fra01}).  Lower-energy electrons with $\gamma <
\gamma_{hr}(\theta_j)$ radiate their energy over a much larger solid
angle.

The mean energy $E_\gamma \equiv m_ec^2 \epsilon =E_j$ of synchrotron
 photons emitted by electrons with $\gamma\simeq 
 \gamma_{hr}(\theta_j)$ is independent of $\psi$ and $B$, and is given
 by
\begin{equation}
 E_j  \cong {\hbar e B\sin\psi\over m_e c}\; 
{ \gamma_{hr}^2(\theta_j) \over  (1+z) }\cong { 500 \over
(\theta_j/ 0.1) [(1+z)/ 2] }\;\;{\rm MeV}\;,
\label{Ej}
\end{equation}
Hyper-relativistic electrons with $\gamma > \gamma_{hr}(\theta_j)$
rapidly lose energy through synchrotron losses and deposit all of
their energy along the direction of the jet. Electrons at lower
energies are deflected to angles $\theta > \theta_j$, and their
emission is not seen by an on-axis observer. Hence the distribution of
electrons along the jet direction always has an effective low-energy
cutoff at $\gamma_{hr}(\theta_j)$. The production spectrum of the
electrons can have an intrinsic cutoff $\gamma_{co}$ due either to the
low-energy cutoff in the escaping neutron spectrum, or to the
neutron-induced photomeson secondary spectrum outside the GRB blast
wave. If $\rho
\equiv\gamma_{co}/ \gamma_{hr}(\theta_j)> 1$, then the observed
synchrotron spectrum is a power law with $-1.5$ index for $
E_j  \lesssim E_\gamma \lesssim \rho^2
 E_j $, and a photon spectrum with the same spectral
index as the accelerated protons and escaping neutrons at photon
energies $E_\gamma \gtrsim \rho^2  E_j $ (\cite{nw91}).
If $\rho < 1$, then the observed photon spectrum at $E_\gamma \gtrsim
E_j $ has the same spectral index as the
primary neutrons.
 
At photon energies $E_\gamma \ll  E_j $, the
observed spectrum is produced by the same hyper-relativistic electrons
with $\gamma \gtrsim \gamma_{hr}(\theta_j)$,
but at energies $\e $ well below the peak energy $ 3\gamma^2 \e_B$, where 
$\e_B \equiv B/B_{cr}$ and $B_{cr} = 4.41\times 10^{13}$ G 
is the critical magnetic field. We now derive this spectrum.

The differential energy radiated per dimensionless energy interval
$d\e$ per differential solid angle element $d\Omega$ in the direction
$\theta$ with respect to the direction of an electron moving with
Lorentz factor $\gamma$ is given by
\begin{equation}
{dE\over d\e d\Omega} = {e^2\over 3\pi^2\lC}
({\e \over \gamma \e_B})^2 (1+\gamma^2\theta^2)^2
(\Lambda_\parallel +
\Lambda_\perp)\;,
\label{dEdedO}
\end{equation}
where $\lC=\hbar/m_ec= 3.86\times 10^{-11}$ cm is the electron Compton
wavelength, and $\Lambda_\parallel = K_{2/3}^2(\xi)$ and
$\Lambda_\perp =
(\gamma\theta)^{2}K_{1/3}^2(\xi)/[1+(\gamma\theta)^{2}]$ are factors
for radiation polarized parallel and perpendicular to the projection
of the magnetic field direction on the plane of the sky defined by the
observer's direction (\cite{jac75}).  The factor $\xi =
\e/\hat\e$, where $\hat \e =
3\e_B\gamma^2/(1+\gamma^2\theta^2)^{3/2}$, and $K_n(x)$ is a modified
Bessel function of the second kind, with asymptotes
$K_n(x) \rightarrow {1\over 2}\Gamma(n )(2/x)^n$ in the limit $x \ll
1$, and $K_n(x) \rightarrow\sqrt{\pi/2x}~\exp(-x)$ in the limit $x \gg
1$. The condition $\xi \ll 1$
corresponds to $\e \ll \hat\e$ where $K_n(\xi )$ are in their 
power-law asymptotes, and $\xi \gtrsim 1$
or $\e \gtrsim \hat \e$ is where $K_n(\xi )$ are in
exponential decline. The characteristic energy $\hat \e$ approaches
 $3\e_B\gamma^2$ when $\gamma\theta \ll 1$, and $\hat
\e$ declines with $\theta$ according to the relation $\hat \e \cong
3\e_B\gamma^2/(\gamma\theta)^3$ when $\gamma\theta \gg 1$.  When
$\e\ll\hat\e$, then $\Lambda_\parallel \gg \Lambda_\perp$ and $dE/d\e
d\Omega \simeq 3^{1/3}(1.07
e/\pi)^2(\gamma\e/\e_B)^{2/3}/\lC \propto \e^{2/3}$.  For a fixed 
value of $\e$, this emissivity
exponentially cuts off  when $\e\gtrsim 3\e_B/\gamma\theta^3$, or when
$\theta \gtrsim \theta_{max}=(3\e_B/\gamma\e)^{1/3}$.

The synchrotron emission spectrum at energies $E_\gamma \ll 
E_j$, integrated over solid angle, is thus
\begin{equation}
{dE\over d\e}\simeq 2\pi \int_0^{(3\e_B/\gamma\e)^{1/3}}d\theta\theta\;
({dE\over d\e d\Omega})\simeq {3e^2\over \pi \lC}\propto \e^0\;.
\label{dEde}
\end{equation}
This differs from the energy index $+1/3$ for synchrotron radiation
radiated by electrons in the classical regime, because in this case
one integrates over a complete orbit of an electron  (\cite{rl79}). 
In this case, the solid angle element $d\Omega \rightarrow
d\theta\sin\theta \rightarrow d\theta \sin\psi $ in the 
integration in eq.\ (\ref{dEde}).

\section{Neutron Beams from GRBs}

In our model, ultrarelativistic protons undergo photomeson
interactions with internal and external radiation photons, producing a
beam of outflowing neutrons.  Subsequent interactions of these
neutrons with photons of the external radiation field generate a beam
of hyper-relativistic electrons.  The $\gtrsim 1$ MeV emission during
the prompt phase is mainly cascade radiation produced within the GRB
shell, while the emission induced by the neutron beam is formed at
later times. This interpretation requires that the fluence of
photohadronic cascade radiation within the shell exceed by a factor of
order unity the cascade fluence induced by the neutral beam radiation,
consistent with observations.  At later times, as shown above,
synchrotron radiation from the hyper-relativistic electrons has a
low-energy cutoff at $E_\gamma < E_j \sim 0.5$ GeV, with a
specific characteristic power-law number spectral index equal to $-1$,
as observed for GRB 941017 (\cite{gon03}).

The $\approx 200$ s decay time of the anomalous emission can be
explained by the projected thickness $\simeq R(1-\cos\theta_j)$ of the
emitting region from the front to the edges of a jet blastwave at
distances $R \approx 6\times 10^{14}(\theta_j/0.1)^{-2}[(1+z)/2]^{-1}$
cm from the central source, implying significant opacity to photomeson
processes due to external photons at these distances. Note that
hyper-relativistic electrons can make a substantial contribution to
emission at $\theta \approx \theta_j\gg 1/\Gamma$, where $\Gamma$ is
the GRB blastwave Lorentz factor.

The external radiation field could be due to plerionic emission in the
supranova model (\cite{kg02}), or to GRB synchrotron photons that are
backscattered by stellar wind electrons in the collapsar model
(\cite{bel02}). We estimate the effective photomeson energy-loss
scattering depth $\tau_{n\gamma}$ for neutrons in the collapsar-model
case, and show that it implies a size scale of the back-scattering
material in accord with the radius inferred from the duration of the
anomalous component. Above the photopion threshold for backscattered
radiation, which applies to $\gtrsim$ several hundred GeV protons,
$\tau_{n\gamma}\approx K_{n\gamma}\sigma_{n\gamma}
\Phi_\gamma (R)\tau_{\rm T}$, where the product of the inelasticity
$K_{n\gamma}$ and cross section $\sigma_{n\gamma}$ is $\cong
70~\mu$barns (\cite{ad03}), $\Phi_\gamma (R)\cong d_L^2 \varphi/[\langle
\e \rangle m_ec^2 R^2]$ is the photon fluence of the prompt GRB
emission at radius $R$, and $\tau_{\rm T}\cong \sigma_{\rm T}R n_w(R)$
is the Thomson depth for scattering photons with observed energies
$\langle \e \rangle m_e c^2$ at radius $R = 10^{14}R_{14}$ cm in the
frame of the stellar wind.  The expression for photon fluence is
written in terms of the energy fluence $\varphi=10^{-4}\varphi_{-4}$
ergs cm$^{-2}$ from a GRB at luminosity distance $d_L=10^{28}d_{28}$
cm.

The stellar wind outflow rate $\dot M = 10^{-4} \dot M_{-4}
M_\odot$/yr and wind speed $v = 2000 v_{2000}$ km s$^{-1}$ are scaled
to the properties of the winds of Wolf-Rayet stars (\cite{loz92}),
which are likely GRB progenitors.  This implies a Thomson depth
$\tau_{\rm T} \cong 0.01 \dot M_{-4}/R_{14} v_{2000}$, so that
$\tau_{n\gamma} \approx d_{28}^2 \varphi_{-4} \dot M_{-4}/[\langle \e
\rangle (1+z) R_{14}^3 v_{2000}]$ at scales $R$.  Equating the 
restriction on radius following from the requirement
$\tau_{n\gamma}\gtrsim 1$ with the relation on $R$ from duration for
$z =1$ and $d_{28} = 2.2$ implies $\theta_j
\cong 0.14 [ (\langle \e \rangle /0.1) v_{2000} /\varphi_{-4} 
\dot M_{-4}]^{1/6}$, which is a
typical jet opening angle.
The differing GRB external radiation and density environments which
determine the intensity of target photons could account for the
unusual spectrum of GRB 941017.  

The fluence in the anomalous MeV component of GRB 941017 is $\gtrsim
3\times$ larger than the prompt emission fluence. If the anomalous
fluence is assumed to originate from photopion processes from
nonthermal protons, then $\gtrsim 1$ order of magnitude more energy
should be available in hadrons than electrons to produce these fluence
ratios, when account is taken of losses to neutrinos and inefficient
extraction of proton energy.  The neutral beam model interpretation
therefore requires that the accelerated particles in, at least, GRB
941017 are hadronically dominated by $\gtrsim 1$ order of magnitude,
consistent with predictions linking cosmic rays to GRB sources
(\cite{wda03}).

The neutron beam can take up to $\approx 30$ -- 50\% of the
accelerated proton energy, with an equal amount of power going into
neutrinos, gamma rays and electrons. At later stages when the internal
photon field disappears, the cascade induced by electrons in the blast
wave is suppressed as a consequence of the strong Klein-Nishina
effects on $\gamma\gamma$ photo-absorption for the external radiation,
which is contributed mostly by backscattered prompt X-rays and MeV
$\gamma$ rays. But the backscattered radiation field still remains
effective for photohadronic processes with UHE protons that continue
to be accelerated in the GRB blast wave after the decline of the
prompt emission. Note that such an environment satisfies requirements
for acceleration by external shocks (\cite{vmg03}), and for
acceleration through the converter mechanism (\cite{der03}). Because
the opacity to $\gamma\gamma$ absorption is small for UHE $\gamma$
rays from $\pi^0$ decay, these $\gamma$ rays will escape from the
blast wave and contribute to the neutral beam power at later stages
directly.

In conjunction with the neutral beam power is the prediction of a very
significant fluence of UHE neutrinos. From GRBs with anomalous
$\gamma$-ray components and large fluences (the fluence of GRB 941017,
including the anomalous component, is $\gtrsim 6.5\times 10^{-4}$ ergs
cm$^{-2}$), we predict several $\nu_\mu$ events detectable with an
IceCube-class km-scale telescope from this type of event, in agreement
with phenomenological inferences by \cite{ahh03}.

In summary, we have proposed a hadronic model for the anomalous
radiation component in GRB 941017.  During the prompt phase, this
emission originates from cascade radiation within the GRB blast
wave. The anomalous component in the extended phase is synchrotron
radiation from hyper-relativistic electrons formed by the associated
neutron beam escaping from the blast wave.  The radiation components
are well-suited for observations with the GBM and LAT on {\it GLAST}.
GRBs with anomalous $\gamma$-ray emission components should also be
bright neutrino sources detectable with {\it IceCube}.

\begin{acknowledgements}
Discussions with M. Gonz\'alez, B.\ Dingus, A.\ K\"onigl and D.\
Lazzati are gratefully acknowledged.  This work is supported by the
Office of Naval Research and the NASA {\it GLAST} program.
\end{acknowledgements}


\begin{thebibliography}{}

\bibitem[Aloisio and Blasi (2002)]{ab02} Aloisio, R., 
and Blasi, P.\ 2002, {\it Astropar. Phys.}, {\bf 18}, 195.

\bibitem[Alvarez-Mu\~niz, et al.\ (2003)]{ahh03} Alvarez-Mu\~niz, J.,
 Halzen, F., \& Hooper, D., {\it Astrophys.\ J.\ Lett.}, 
 submitted (astro-ph/0310417) 

\bibitem[Atoyan and Dermer (2003)]{ad03} Atoyan, A., 
and Dermer, C.\ D.\ 2003, {\it Astrophys.\ J.}, {\bf 586}, 79.

\bibitem[Beloborodov (2002)]{bel02} Beloborodov, A.\ M.\ 2002,
{\it Astrophys.\ J.}, {\bf 565}, 808. 

\bibitem[Derishev, et al.\ (2003)]{der03} Derishev, E.~V., 
Aharonian, F.~A., Kocharovsky, V.~V., \& Kocharovsky, V.~Vl.\ 2003, 
{\it Phys.\ Rev.\ D}, {\bf 68}, 43003 

\bibitem[Dermer (2002)]{der02} Dermer, C.\ D.\ 2002, 
{\it Astrophys.\ J.}, {\bf 574}, 65.

\bibitem[Dermer and Atoyan (2003)]{da03} Dermer, C.\ D., 
and Atoyan, A.\ 2003, {\it Phys.\ Rev.\ Lett.}, 
{\bf 91}, 071102.

\bibitem[Frail et al.\ (2001)]{fra01} Frail, D., et al.\ 2001, 
{\it Astrophys.\ J.}, {\bf 562}, L55.

\bibitem[Gonz\'alez et al.\ (2003)]{gon03} Gonz\'alez,
 M.\ M., Dingus, B.\ L., Kaneko, Y., et al.\ 2003, 
{\it Nature}, {\bf 424}, 749.

\bibitem[Granot and Guetta (2003)]{gg03} Granot, J., 
and Guetta, D.\ 2003, {\it Astrophys.\ J.\ Lett.}, 598, L11.

\bibitem[Jackson (1975)] {jac75} Jackson, J.\ D. 1975, 
{\it Classical Electrodynamics} (Wiley, New York).     

\bibitem[K\"onigl and Granot (2002)]{kg02}K\"onigl, A., and Granot,
 J.\ 2002, {\it Astrophys.\ J.}, {\bf 574}, 134.

\bibitem[Lozinskaya (1992)]{loz92} Lozinskaya, T. 1992, {\it Supernovae
and Stellar Wind in the Interstellar Medium} (New York: AIP).

\bibitem[Nelson and Wasserman (1991)]{nw91} Nelson,
 R.\ W., and Wasserman, I.\ 1991, {\it Astrophys.\ J.}, {\bf 371}, 265.

\bibitem[Pe'er and Waxman (2004)] {pw04} Pe'er, A., and Waxman, E.\ 2004, 
 {\it Astrophys.\ J.\ Lett.}, {\bf 603}, L1.

\bibitem[Rachen and M{\' e}sz{\' a}ros (1998)]{rm98} Rachen, 
J.~P., and M{\' e}sz{\' a}ros, P.\ 1998, {\it Phys.\ Rev.\ D.}, 
{\bf 58}, 123005. 

\bibitem[Rybicki and Lightman (1979)] 
{rl79} Rybicki, G.\ B., and Lightman, A.\ P.\ 1979, 
{\it Radiative Processes in Astrophysics} (Wiley, New York).

\bibitem[Vietri (1995)]{vie95} Vietri, M. 1995, 
{\it Astrophys.\ J.}, {\bf 453}, 883.

\bibitem[Vietri, De Marco, \& Guetta (2003)]{vmg03} Vietri,
M., De Marco, D., \& Guetta, D.\ 2003, {\it Astrophys.\ J.}, 
{\bf 592}, 378

\bibitem[Waxman (1995)]{wax95} Waxman, E. 1995, 
{\it Phys.\ Rev. Lett.}, {\bf 75}, 386.

\bibitem[Wick et al.\ (2003)]{wda03}Wick, S.\ D., 
Dermer, C.\ D., and Atoyan, A.\ 2003, 
{\it Astropar.\ Phys.}, in press (astro-ph/0310667). 

\end{thebibliography}
\end{document}